\documentstyle[amsfonts,amssymb]{article}    

\newcommand{\mb}{\mathbb}

\newcommand{\be}{\begin{equation}}
\newcommand{\ee}{\end{equation}}
\newcommand{\beq}{\begin{eqnarray}}
\newcommand{\eeq}{\end{eqnarray}}
\newcommand{\beqst}{\begin{eqnarray*}}
\newcommand{\eeqst}{\end{eqnarray*}}

\newcommand{\pa}{\partial}

\newcommand{\dsp}{\displaystyle}

\newtheorem{theorem}{Theorem}[section]
\newtheorem{lemma}[theorem]{Lemma} 
 
\newtheorem{corollary}[theorem]{Corollary} 
\newtheorem{remark}[theorem]{Remark} 
\newtheorem{proposition}[theorem]{Proposition}

\title{The semilinear Klein-Gordon equation \\ in  de~Sitter spacetime}

\author{{\bf Karen Yagdjian\thanks{Correspondence:  Department of Mathematics, 
University of Texas-Pan American 
1201 W. University Drive, 
Edinburg, TX 78541-2999, 
USA; E-mail:yagdjian@utpa.edu. }}}

\date{}
\begin{document}
\maketitle

\bigskip

\begin{abstract}
 In this article we study  the blow-up phenomena for the solutions of the semilinear 
Klein-Gordon equation $\Box_g \phi-m^2 \phi  = -|\phi |^p  $ with the small mass $m \le n/2$ in  de~Sitter space-time with the metric $g$. 
We prove that for every $p>1$ the large energy solution blows up, while for the small energy solutions we give a borderline $p=p(m,n)$
for the global in time existence. The consideration is based on the representation formulas for the solution of the Cauchy problem and 
on some generalizations of the Kato's lemma.
\end{abstract}

\section{Introduction}

 In this article we study  the blow-up phenomena for the solutions of the semilinear 
Klein-Gordon equation $\Box_g \phi-m^2 \phi  = -|\phi |^p  $ with the small mass $m \le n/2$  in  de~Sitter space-time.  
\smallskip

In the model  of the universe proposed by  de~Sitter  the line
element  has the   
form
\[
ds^2=-\left( 1-\frac{2M_{bh}}{r} - \frac{\Lambda r^2}{3}\right)c^2dt^2 + \left( 1-\frac{2M_{bh}}{r} - \frac{\Lambda r^2}{3}\right)^{-1}dr^2 + r^2(d \theta ^2+ \sin^2 d\phi ^2).
\]
The constant  $M_{bh}$ may have a meaning of the ``mass of the black hole''. 
The corresponding metric with this line element 
is called the Schwarzschild - de~Sitter metric. 
\smallskip

The Cauchy problem for the semilinear Klein-Gordon equation   in  Minkowski spacetime ($M_{bh}=\Lambda =0 $) is well investigated. (See, e.g., \cite{Keel-Tao} and 
references therein.)
In particular, Keel and Tao~\cite{Keel-Tao} for the semilinear equation 
$u_{tt}-\Delta u=F(u )$, $u(0,x)=\varepsilon \varphi_0 (x)$, $u_t(0,x)=\varepsilon \varphi _1 (x)$ proved that if $n = 1, 2, 3$
 and $1 < p < 1 + 2n$, then
there exists  a (non-Hamiltonian) nonlinearity 
$F$ satisfying $|D^\alpha F(u)|\le C|u|^{p-|\alpha |}$ for $0\le \alpha \le [p]$ and such that there is no finite energy global solution  
 supported in the forward light cone, for any nontrivial smooth compactly supported $ \varphi_0$ and $ \varphi  _1$ and for any $\varepsilon >0$. 
There is an interesting question of instability of the ground state standing solutions $e^{i\omega t}\phi_\omega(x)$ for nonlinear Klein-Gordon equation  
 $\partial_t^2u-\Delta u+u=|u|^{p-1}u$. Here  $\phi _\omega $ is a ground state of 
the equation $ -\Delta \phi +(1-\omega ^2)\phi = |\phi |^{p-1}\phi,$  
while  $0<p-1< 4/(N-2)$ and $0\le |\omega |<1$.
Ohta and Todorova~\cite{Ohta-Todorova}  showed that instability occurs in the very strong 
sense that an arbitrarily small perturbation of the initial data can make the perturbed 
solution blow up in finite time.    
 \smallskip

The Cauchy problem for the linear wave equation without source term   on  
the maximally extended 
Schwarzschild - de Sitter spacetime in the case of non-extremal black-hole corresponding to parameter values
$0<M_{bh}< \frac{1}{3\sqrt{\Lambda }}$,  
is considered  by Dafermos and Rodnianski~\cite{Dafermos-Rodnianski_1}.  
They   proved that in the region bounded by a set of black/white hole horizons and cosmological horizons, 
solutions converge pointwise to a constant faster than any given polynomial rate, 
where the decay is measured with respect to natural future-directed advanced and retarded time coordinates.
The bounds on decay rates for solutions to the wave equation in the Schwarzschild - de Sitter spacetime is a first step 
to a mathematical understanding of non-linear stability problems for spacetimes containing black holes.
\smallskip

Catania and Georgiev~\cite{Catania-Georgiev} studied  the Cauchy problem for the semilinear 
wave equation $\Box_g \phi = |\phi |^p  $ in the Schwarzschild metric $(3+1)$-dimensional space-time, that is the  case of $\Lambda =0 $
in $0<M_{bh}< \frac{1}{3\sqrt{\Lambda }}$. 
They established that the problem in the Regge-Wheeler coordinates  is locally well-posed in $H^\sigma$ for any $\sigma\in[1,p+1)$. Then for the 
special choice of the initial data they proved the blow-up of the solution in two cases: 
(a) $p\in(1,1+\sqrt2)$ and small initial data supported far away from the black hole; (b) $p\in(2,1+\sqrt2)$ and large data supported near the 
black hole. In both cases, they also gave an estimate from above for the lifespan of the solution.

\smallskip

In the present paper we focus on the another limit case as  \,$M_{bh} \to 0$ in $0<M_{bh}< \frac{1}{3\sqrt{\Lambda }}$, namely, we  set \,$M_{bh}=0$ \,to ignore completely influence of the black hole.  
Thus,  the  line element in de~Sitter spacetime has  the form
\[
ds^2= - \left( 1- \frac{r^2}{R^2}\right) c^2\, dt^2+ \left( 1- \frac{r^2}{R^2}\right)^{-1}dr^2 + r^2(d\theta ^2 + \sin^2 \theta \, d\phi ^2)\,.
\]
The  Lama{\^i}tre-Robertson transformation \cite{Moller} 
\[
r'=\frac{r}{\sqrt{1-r^2/R^2}} e^{-ct/R}\,, \quad t'=t+\frac{R}{2c} \ln \left( 1- \frac{r^2}{R^2}\right) \,, \quad \theta '=\theta \,, 
\quad \phi '=\phi, 
\]
leads to the following form for the line element: $
ds^2= -   c^2\, d{t'}^2+e^{2ct'/R}( d{r'}^2   + r'^2\,d{\theta'} ^2 + r'^2\sin^2 \theta' \, d{\phi '}^2)$. 
By defining  coordinates $x'$, $y'$, $z'$ connected with $r'$, $\theta '$, $\phi '$
by the usual equations connecting Cartesian  coordinates and polar coordinates in a Euclidean space, the line element may be written 
\cite[Sec.134]{Moller}
\[
ds^2= -   c^2\, d{t'}^2+ e^{2ct'/R}( d{x'}^2   + d{y'} ^2 +  d{z '}^2)\,.
\]
The new coordinates $r'$, $\theta '$, $\phi '$, $t'$ can take all values from $-\infty$ to $\infty$.
Here $R$ is the ``radius'' of the universe.

\smallskip

In this paper we  study blow-up phenomena for semilinear equation by applying the Lama{\^i}tre-Robertson 
 transformation and by employing the fundamental solutions  for some 
model linear
hyperbolic equation with variable speed of propagation.  
 In  \cite{Yag_Galst_CMP} the   Klein-Gordon  operator in Robertson-Walker spacetime, that is 
${\mathcal S}:= \partial_t^2  - e^{-2t}\bigtriangleup  + M^2 $,
is considered. The  fundamental solution  $E=E(x,t;x_0,t_0)$, that is solution of \,${\mathcal S} E = \delta (x-x_0,t-t_0)$, 
with a support in the forward light cone $D_+ (x_0,t_0) $, $x_0 \in {\mathbb R}^n$, $t_0 \in {\mathbb R}$,
and  the  fundamental solution with a support in the backward light 
cone $D_- (x_0,t_0) $, $x_0 \in {\mathbb R}^n$, $t_0 \in {\mathbb R}$,
defined by $ 
D_\pm (x_0,t_0) 
 := 
\big\{ (x,t)  \in {\mathbb R}^{n+1}  \, ; \, 
|x -x_0 | $ $\leq \pm( e^{-t_0} - e^{-t })
\,\big\} $,  
are constructed.
These fundamental solutions have been used to  represent     solutions of the Cauchy problem and  to prove   $L^p-L^q$ estimates 
for the solutions of the equation with and without a source term
 that provide with some necessary tools for the studying semilinear equations.

\smallskip

In the Robertson-Walker spacetime \cite{Hawking}, one can choose coordinates so that the metric has the form
\[
ds^2=-dt^2+S^2(t)d \sigma ^2\,.
\]
In particular, the  metric in 
de Sitter and anti-de Sitter  spacetime in the Lama{\^i}tre-Robertson coordinates \cite{Moller} has this form with $S (t)=e^{t} $
and  $S (t)=e^{-t} $, respectively. 
The matter waves in the de~Sitter spacetime are described by the function  $\phi $, which  satisfies equations of 
motion.  
In the  de~Sitter universe  the equation for the scalar field with mass \,  $m$\,   and potential function \, $V$   \, 
is the covariant Klein-Gordon equation  
\[
\square_g \phi  - m^2 \phi  = V'(\phi ) \quad \mbox{\rm or} \quad \frac{1}{\sqrt{|g|}}\frac{\partial }{\partial x^i}
\left( \sqrt{|g|} g^{ik} \frac{\partial \phi  }{\partial x^k} \right)- m^2\phi  =V'(\phi ) \,,
\]
with the usual summation convention. Written explicitly in coordinates  in the  de Sitter spacetime it, in particular, 
for \,$V'(\phi )=  -|\phi |^p$ \,  has  the form  
\begin{equation}
\label{1.1}
  \phi_{tt} +   n   \phi_t - e^{-2 t} \Delta  \phi + m^2\phi=    |\phi |^p\,.
\end{equation}
In this paper we restrict ourselves with consideration of the semilinear equation for particle with small mass \,$m$, 
that is \,$0 \leq m \leq n/2 $. If we introduce the new unknown function \,$u = e^{\frac{n}{2}t}\phi$, then it takes the form of the semilinier 
Klein-Gordon  equation for \,$u$\, on de~Sitter spacetime 
\begin{equation}
\label{2.2}
u_{tt} - e^{-2t} \bigtriangleup u  - M^2 u=   e^{-\frac{n(p-1)}{2}t}|u |^p,
\end{equation}
where non-negative curved mass \,$M \geq 0$\, is defined as follows:
\[
M^2:=  \frac{n^2}{4} - m^2\geq 0\,.
\]
The equation (\ref{2.2}) can be regarded as Klein-Gordon  equation with imaginary mass. Equations with imaginary mass appear in 
several physical  models such as \, $\phi ^4$ \, field model, tachion (super-light) fields,  Landau-Ginzburg-Higgs equation  and others.
 To solve the Cauchy problem for semilinear equation we use fundamental solution of the corresponding linear operator. 
We denote by $G$ the resolving operator of the problem 
\begin{equation}
\label{2.2a}
u_{tt} - e^{-2t} \bigtriangleup u  - M^2 u=   f,   \quad u  (x,0) =  0 , \quad \partial_{t }u  (x,0 ) =0\,.
\end{equation}
Thus, $u=G[f]$. The  equation of (\ref{2.2a}) is strictly hyperbolic. This implies the well-posedness of the Cauchy problem (\ref{2.2a})  
in the different functional spaces. Consequently, the operator is well-defined in those  functional spaces.
\smallskip

Then, the speed of propagation is variable,  namely, 
it is equal to $e^{-t} $. 
The second-order strictly hyperbolic  equation (\ref{2.2a}) possesses two fundamental solutions 
resolving the Cauchy problem without source term $f$. They can be written in terms of the Fourier integral operators, which
give complete description of the wave front sets of the solutions.  
Moreover, the  integrability
of the characteristic roots, $\int_0^{\infty} |\lambda_i (t,\xi )| dt <  \infty $, $i=1,2$, leads to the 
existence of the so-called ``horizon'' for that equation. More precisely,  any  signal emitted from the spatial point $x_0 \in {\mathbb R}^n$
at time $t_0 \in {\mathbb R} $ remains inside the ball $B^n_{t_0}(x_0):= \{x \in {\mathbb R}^n\,|\,|x -x_0 | <e^{-t_0} \}$ for all time $t \in (t_0 ,\infty ) $. 
In particular, it can cause a nonexistence of the $L^p-L^q$ decay 
for the solutions. In \cite{yagdjian_birk}  this phenomenon is illustrated 
by a model equation with permanently bounded
domain of influence, power decay of characteristic roots, and without 
$L^p-L^q$ decay. The above mentioned $L^p-L^q$ decay 
estimates are one of the important tools for studying nonlinear problems
(see, e.g. \cite{Shatah}). In this paper we show that this phenomenon  causes the blow up of the solution.
The  equation (\ref{2.2a})
is neither Lorentz  invariant nor  invariant with respect to usual scaling and that creates additional difficulties. 

\smallskip

Operator $G$ is constructed in \cite{Yag_Galst_CMP} for the case of the large mass\, $m  \geq n /2$. 
The analytic continuation of this operator in parameter $M $ into ${\mathbb C}$  allows us to use $G$
 also in the case of small mass\, $0 \leq m  \leq n /2$.  More precisely, 
 we define the operator \,$G$\, acting on \,$f(x,t) \in C^\infty ({\mb R}\times [0,\infty)) $ \,by 
\begin{eqnarray*}
&  &
G[f](x,t)  \\
& := &
 \int_{ 0}^{t} db \int_{ x - (e^{-b}- e^{-t})}^{x+ e^{-b}- e^{-t}}  dy\, f(y,b)  
(4e^{-b-t })^{-M} \Big((e^{-t }+e^{-b})^2 - (x - y)^2\Big)^{-\frac{1}{2}+M    } \\
&  &
\hspace{2.5cm}\times F\Big(\frac{1}{2}-M   ,\frac{1}{2}-M  ;1; 
\frac{ ( e^{-b}-e^{-t })^2 -(x- y )^2 }{( e^{-b}+e^{-t })^2 -(x- y )^2 } \Big)   , 
\end{eqnarray*}
where $F\big(a, b;c; \zeta \big) $ is the hypergeometric function.(See, e.g., \cite{B-E}. For analytic continuation see , e.g., 
 \cite[Sec. 1.8]{Slater} .) If $n$ is odd, $n=2m+1$,  $m  \in {\mathbb N}$,  then for $f\in C^\infty  ({\mb R}^n\times [0,\infty)) $, we define
\begin{eqnarray*}
 & &
 G[f](x,t)  \\
& := &
2\int_{ 0}^{t} db
  \int_{ 0}^{ e^{-b}- e^{-t}} dr_1 \,  \left(  \frac{\partial }{\partial r} 
\Big(  \frac{1}{r} \frac{\partial }{\partial r}\Big)^{\frac{n-3}{2} } 
\frac{r^{n-2}}{\omega_{n-1} c_0^{(n)} }  \!\!\int_{S^{n-1} } f(x+ry,b)\, dS_y  
\right)_{r=r_1}   \nonumber \\
  &  & 
\hspace{2.5cm}\times  (4e^{-b-t})^{-M}
\left( (e^{-t}  + e^{-b} )^2 - r_1^2   \right)^{-\frac{1}{2}+M}\\
  &  & 
\hspace{2.5cm}\times 
F\left(\frac{1}{2}-M,\frac{1}{2}-M;1; 
\frac{ (e^{-b}- e^{-t})^2-r_1^2}
{  (e^{-b}+ e^{-t})^2-r_1^2} \right)\!\! ,
\end{eqnarray*} 
where $c_0^{(n)} =1\cdot 3\cdot \ldots \cdot (n-2 )$.
Constant $\omega_{n-1} $ is the area of the unit sphere $S^{n-1} \subset {\mathbb R}^n$. 

If $n$ is even, $n=2m$,  $m  \in {\mathbb N}$,  then for $f  \in C^\infty   ({\mb R}^n\times [0,\infty)) $,  
the operator $G$  is given by the next expression 
\begin{eqnarray*}
\hspace{-0.3cm} &  &
G[f](x,t)  \\
& := &
2\int_{ 0}^{t} db
  \int_{ 0}^{ e^{-b}- e^{-t}} dr_1 \,  \left( \frac{\partial }{\partial r} 
\Big( \frac{1}{r} \frac{\partial }{\partial r}\Big)^{\frac{n-2}{2} } 
\frac{2r^{n-1}}{\omega_{n-1} c_0^{(n)} }  \!\!\int_{B_1^{n}(0)} \frac{f(x+ry,b) }{\sqrt{1-|y|^2}} \, dV_y 
\right)_{r=r_1}   \nonumber \\
  &  & 
\hspace{2.5cm}\times  (4e^{-b-t})^{-M}
\left( (e^{-t}  + e^{-b} )^2 - r_1^2   \right)^{-\frac{1}{2}+M}\\
  &  & 
\hspace{2.5cm}
\times F\left(\frac{1}{2}-M,\frac{1}{2}-M;1; 
\frac{ (e^{-b}- e^{-t})^2-r_1^2}
{  (e^{-b}+ e^{-t})^2-r_1^2} \right)\!\! .
\end{eqnarray*} 
Here $B_1^{n}(0) :=\{|y|\leq 1\} $ is the unit ball in ${\mathbb R}^n$, while $c_0^{(n)} =1\cdot 3\cdot \ldots \cdot (n-1)$.
Thus, in both cases, of even and odd $n$, one can write 
\begin{eqnarray*} 
u(x,t) 
&  =  &
2   \int_{ 0}^{t} db
  \int_{ 0}^{ e^{-b}- e^{-t}} dr  \,  v(x,r ;b)  (4e^{-b-t})^{-M}
\left( (e^{-t}  + e^{-b} )^2 - r^2   \right)^{-\frac{1}{2}+M}  \nonumber \\
&  &
\hspace{2.5cm} 
\times F\left(\frac{1}{2}-M,\frac{1}{2}-M;1; 
\frac{ (e^{-b}- e^{-t})^2-r^2}
{  (e^{-b}+ e^{-t})^2-r^2} \right) ,
\end{eqnarray*}
where the function 
$v(x,t;b)$   
is a solution to the Cauchy problem for the  wave equation
\[
v_{tt} -   \bigtriangleup v  =  0 \,, \quad v(x,0;b)=f(x,b)\,, \quad v_t(x,0;b)= 0\,.
\]
It can be proved   that if $ n\big( 1 - \frac{2}{q} \big) \leq 1$, then for every given $T>0$
the operator $G$ can be extended to the bounded operator:
\[
G \,: \, C([0,T]; L^{q'}({\mb R}^n)) \longrightarrow  C([0,T]; L^q({\mb R}^n)) \,.
\]
Consequently the operator $G$ maps
\[
G \,: \, C([0,\infty); L^{q'}({\mb R}^n)) \longrightarrow  C([0,\infty); L^q({\mb R}^n)) ,
\]
in the corresponding topologies.  Moreover, 
\[
G \,: \, C([0,\infty); L^{q'}({\mb R}^n)) \longrightarrow  C^1([0,\infty); {\mathcal D}'({\mb R}^n))  .
\]

Let $u_0=u_0(x,t)$ be a solution of the Cauchy problem
\begin{equation}
\label{3.5}
\pa_{t }^2 u_0 - e^{-2t} \bigtriangleup u_0   - M^2 u_0 =   0, \quad u_0 (x,0) = \varphi _0(x), \quad \pa_{t }u_0 (x,0 ) = \varphi _1(x )\,.
\end{equation}
Then any solution $u=u(x,t)$ of the equation (\ref{2.2}) which takes initial value $ u (x,0) = \varphi _0(x), \quad \pa_{t }u (x,0) = \varphi _1(x)$,  
solves also integral equation
\begin{equation}
\label{2.6new}
u(x,t) = u_0(x,t)+ G[e^{-\frac{n(p-1)}{2}\cdot }|u |^p](x,t)   \,.
\end{equation}
Let $ \Gamma \in C([0,\infty))$. For every given function $u_0 \in C([0,T]; L^{q'}({\mb R}^n))$ we consider integral equation (\ref{2.6new}) 
\begin{equation}
\label{2.7}
u(x,t) = u_0(x,t)+ G\left[\Gamma (\cdot ) \left( \int_{{\mb R}^n} |u (y,\cdot )|^p dy \right)^\beta  |u (y,\cdot )|^p\right](x,t)    \,,
\end{equation}
for the function $u  \in C([0,T]; L^q({\mb R}^n))\cap C([0,T]; L^p({\mb R}^n))$. 
Here $ q' \geq  q>1$, $p  \geq 1$. The last integral equation 
corresponds to the slightly more general equation than (\ref{2.2}), namely, to the {\it nonlocal} equation
\begin{equation}
\label{2.8}
u_{tt} - e^{-2t} \bigtriangleup u  - M^2 u=   \Gamma (t ) \left( \int_{{\mb R}^n} |u (y,t )|^p dy \right)^\beta |u |^p \,. 
\end{equation}
 If $u_0$ is generated by the Cauchy problem (\ref{3.5}), then the solution  $u=u (x,t)$ of (\ref{2.7}) 
is said to be  
{\it a weak solution} of the Cauchy problem with the initial conditions
\[ 
u (x,0) = \varphi _0 (x), \quad \pa_{t }u (x,0) = \varphi _1 (x)\,,
\]
for the equation (\ref{2.8}). In the present paper we are looking for the conditions on the function $\Gamma $, on constants $M$,  $n$, $p$,
 and $ \beta $ that guarantee   a non-existence of global in time weak solution, namely, the blow-up phenomena. 
We are especially interested in the scale of functions $\Gamma (t)=(1+t)^{d_1}e^{d_0t}$, where $d_0, d_1 \in {\mathbb R}$.
The function $\,e^{-\frac{n(p-1)}{2}t}$ \,with  $d_0= -n(p-1)/2$  and $d_1=0 $ is in 
that scale and represents equation (\ref{2.2}) if $ \beta =0$. In particular, we   
find in the next theorem the upper  bound for $d_0$  with an existence of 
the global solution for small initial data. For   equation (\ref{2.8}) in that scale the bound 
is given by \, $d_0 \geq  -M(p(\beta +1)-1)$\, and\, $ d_1>2$ \, if \,$M>0$.

\smallskip

\begin{theorem}
\label{T2.1}
Suppose that function  $ \Gamma \in C^1([0,\infty))$ is either non-decreasing or non-increasing, and if $M>0$ then 
\beqst
&  &
 \Gamma (t) \geq c e^{-M(p(\beta +1)-1 )t} t^{2+\varepsilon}\quad \mbox{\rm for all} \quad t \in [0,\infty) ,
\eeqst
with the numbers $\varepsilon >0$ and $ c>0$, while for $M=0$ it satisfies
\beqst
&  &
\Gamma (t) \geq c t^{-1-p(\beta +1)} \,.
\eeqst

Then, for every $p>1 $, $N$, and $\varepsilon  $ there exists $u_0 \in C^\infty({\mb R}^n \times [0,\infty))$  which 
for any given slice of constant time   $t =const \ge 0$ has a compact support in $x$, 
such that $u_0 (x,0), \partial_t u_0 (x,0) \in C^\infty_0(  {\mb R}^n )$, and
\[
\|u_0 (x,0)\|_{C^N({\mathbb R}^n)} + \|\partial_t u_0 (x,0)\|_{C^N({\mathbb R}^n)} < \varepsilon 
\]
but  a global in time solution $u  \in C([0,\infty); L^q({\mb R}^n))$ of the equation (\ref{2.7}) with permanently bounded support
does not exist for all $q \in[2, \infty)$  
and $ \beta >1/p-1 $. More precisely,
there is $T>0$ such that 
\[
\lim_{t  \nearrow T } \int_{{\mb R}^n} u (x ,t) dx  = \infty \,.
\]
\end{theorem}
This theorem shows that instability of the trivial solution occurs in the very strong 
sense, that is,  an arbitrarily small perturbation of the initial data can make the perturbed 
solution blowing up in finite time.   
\smallskip
 
If we allow large initial data, then according to the next theorem, for every  $d_0 \in {\mathbb R}$ and $M>0$ the solution blows up in finite time.   
\begin{theorem}
\label{T2.1a}
Suppose that function  $ \Gamma (t)= e^{\gamma t}$, where $\gamma \in {\mathbb R}$   and that the curved mass is positive,  $M>0$.  
Then, for every \,$p>1$ \, and $n$ there exists $u_0 \in C^\infty( {\mb R}^n \times[0,\infty))$  which 
for any given slice of constant time   $t =const \ge 0$ has a compact support in $x$,   such that $u_0 (x,0), \partial_t u_0 (x,0) \in C^\infty_0(  {\mb R}^n )$ 
but  a global in time solution $u  \in C([0,\infty); L^q({\mb R}^n))$ of the equation (\ref{2.7}) with permanently bounded support
does not exist for all $q \in[2, \infty)$ 
and \,$ \beta >  1/ p -1 $. More precisely,
there is \,$T>0$ \,such that 
\[
\lim_{t  \nearrow T } \int_{{\mb R}^n} u (x ,t) \, dx = \infty \,.
\]
\end{theorem}

Thus, for every $p>1$ the large energy classical solution of the Cauchy for equation (\ref{1.1}) blows up. We will prove  global existence of the small
energy solution   in a forthcoming paper.   
\smallskip

The remaining part of this paper is organized as follows. In Section~\ref{S_sinh} we prove some auxiliary 
integral representations for the function $\sinh(t)$ and the linear function via Gauss's hypergeometric function 
and multidimensional integrals involving also fundamental solution of the Cauchy problem for wave equation in Minkowski spacetime.
 In Section~\ref{S_diffineq} we suggest two simple 
generalizations of Kato's lemma, which allow  us to handle the case  of differential inequalities  with exponentially decreasing kernels. 
In Section~\ref{S_Proofs} we complete the proofs of both theorems.

\section{Integral representations of function $ M^{-1} \sinh (M(t-b)) $ involving hypergeometric function} 
\label{S_sinh}

In \cite[Sec. 2.4]{B-E} one can find one-dimensional integrals involving hypergeometric function. In this section we
present one more example of such integral and also examples of multidimensional integrals appearing in the fundamental solutions
for the Klein-Gordon equation in de~Sitter spacetime. More examples related to the Tricomi and Gellerstedt equations one can find in
\cite{YagTricomi_GE}. 
\begin{proposition} 
\label{Lt_tau}
The function $M^{-1}\sinh (M(t-b))$ with $t\ge b \ge 0$, 
can be represented as follows:\\
\mbox{\rm (i)} The one-dimensional integral  
\begin{eqnarray} 
\label{3.9}
 \frac{1}{M} \sinh (M(t-b ) )   
& = &
  \int_{  - (e^{-b}- e^{-t})}^{e^{-b}- e^{-t}} 
(4e^{-b-t })^{-M} \Big((e^{-t }+e^{-b})^2 - z^2\Big)^{-\frac{1}{2}+M    }   \\
&   &
\times F\Big(\frac{1}{2}-M   ,\frac{1}{2}-M  ;1; 
\frac{ ( e^{-b}-e^{-t })^2 -z^2 }{( e^{-b}+e^{-t })^2 -z^2 } \Big) \,   dz ; \nonumber
\end{eqnarray}
\mbox{\rm (ii)} If $n$ is odd,  $n=2m+1$, $m  \in {\mathbb N}$, then with $c_0^{(n)} =1\cdot 3\cdot \ldots \cdot (n-2 )$, 
\begin{eqnarray*}
& &
 \frac{1}{M} \sinh (M(t-b ) )  \\ 
& = &
2
  \int_{ 0}^{ e^{-b}- e^{-t}} dr_1 \,  \left(  \frac{\partial }{\partial r} 
\Big(  \frac{1}{r} \frac{\partial }{\partial r}\Big)^{\frac{n-3}{2} } 
\frac{r^{n-2}}{\omega_{n-1} c_0^{(n)} }  \!\!\int_{S^{n-1} } \, dS_y  
\right)_{r=r_1}  \!\! (4e^{-b-t})^{-M} \nonumber \\
  &  & 
\times  
\left( (e^{-t}  + e^{-b} )^2 - r_1^2   \right)^{-\frac{1}{2}+M}
F\left(\frac{1}{2}-M,\frac{1}{2}-M;1; 
\frac{ (e^{-b}- e^{-t})^2-r_1^2}
{  (e^{-b}+ e^{-t})^2-r_1^2} \right); 
\end{eqnarray*} 
\mbox{\rm (iii)} If $n$ is even,  $n=2m$, $m  \in {\mathbb N}$, 
then with $c_0^{(n)} =1\cdot 3\cdot \ldots \cdot (n-1 )$, 
\begin{eqnarray*}
& &
 \frac{1}{M} \sinh (M(t-b ) )  \\
& = &
2\int_{ 0}^{ e^{-b}- e^{-t}}  \!\!dr_1  \!\! \left( \frac{\partial }{\partial r} 
\Big( \frac{1}{r} \frac{\partial }{\partial r}\Big)^{\frac{n-2}{2} } 
\frac{2r^{n-1}}{\omega_{n-1} c_0^{(n)} }  \!\!\int_{B_1^{n}(0)} \frac{1 }{\sqrt{1-|y|^2}}  dV_y 
\right)_{r=r_1}  \nonumber \\
  &  & 
\hspace{2.5cm} \times 
 (4e^{-b-t})^{-M} \left( (e^{-t}  + e^{-b} )^2 - r_1^2   \right)^{-\frac{1}{2}+M}\\
  &  & 
\hspace{2.5cm} \times 
F\left(\frac{1}{2}-M,\frac{1}{2}-M;1; 
\frac{ (e^{-b}- e^{-t})^2-r_1^2}
{  (e^{-b}+ e^{-t})^2-r_1^2} \right)\!\! . 
\end{eqnarray*} 
Here 
the constant $\omega_{n-1} $ is the area of the unit sphere $S^{n-1} \subset {\mathbb R}^n$. 
\end{proposition}
\medskip

\noindent
{\bf Proof.}  First we consider case (i). According to Theorem~0.3~\cite{Yag_Galst_CMP} 
for every  function $f \in $ $C^\infty ({\mathbb R}\times [0,\infty))$, which 
for any given slice of constant time   $t =const \ge 0$ has a compact support in $x$, the function
\begin{eqnarray*}
v(x,t)  & = &
 \int_{ 0}^{t} db \int_{ x - (e^{-b}- e^{-t})}^{x+ e^{-b}- e^{-t}}  dy\,
(4e^{-b-t })^{-M} \Big((e^{-t }+e^{-b})^2 - (x - y)^2\Big)^{-\frac{1}{2}+M    } \\
&  &
\hspace{2.3cm}\times F\Big(\frac{1}{2}-M   ,\frac{1}{2}-M  ;1; 
\frac{ ( e^{-b}-e^{-t })^2 -(x- y )^2 }{( e^{-b}+e^{-t })^2 -(x- y )^2 } \Big) f(y,b)   
\end{eqnarray*}
is a unique  $C^\infty $-solution to the Cauchy problem  
\begin{equation}
\label{3.10}
\pa_{t }^2v - e^{-2t} \bigtriangleup v   - M^2 v =   f, \quad v (x,0) = 0, \quad \pa_{t }v (x,0) = 0
\end{equation}
with $n=1$.
 It follows
\begin{eqnarray}
\label{3.11}
&  &
\int_{-\infty}^\infty  v(x,t) dx \\
& = &
 \int_{ 0}^{t} db \Big( \int_{-\infty}^\infty  f(x,b) dx \Big)  
\int_{   - (e^{-b}- e^{-t})}^{  e^{-b}- e^{-t}}  dz    
(4e^{-b-t })^{-M} \Big((e^{-t }+e^{-b})^2 - z^2\Big)^{-\frac{1}{2}+M    } \nonumber \\
&  &
\hspace{2.5cm}\times F\Big(\frac{1}{2}-M   ,\frac{1}{2}-M  ;1; 
\frac{ ( e^{-b}-e^{-t })^2 -z^2 }{( e^{-b}+e^{-t })^2 -z^2 } \Big)  \nonumber   .
\end{eqnarray}
On the other hand, from the linear Klein-Gordon equation (\ref{3.10})  and the vanishing initial data, we obtain
\begin{eqnarray*}
&  &
\hspace*{-0.0cm} \int_{-\infty}^\infty  v(x,t) dx - M^2 \int_0^t d\tau \int_0^\tau  db \int_{-\infty}^\infty  v(x,b) dx \\
& = &
\int_0^t d\tau \int_0^\tau  db \int_{-\infty}^\infty e^{-2b}  \partial_x^2 v(x,b) dx +
\int_0^t d\tau \int_0^\tau db \int_{-\infty}^\infty  f(x,b)\,dx   \,,
\end{eqnarray*}
that is
\begin{eqnarray}
\label{1.8}
&  & 
\int_{-\infty}^\infty  v(x,t) dx - M^2 \int_0^t d\tau \int_0^\tau  db \int_{-\infty}^\infty  v(x,b) dx \\
& = &
\int_0^t \Big( \int_{-\infty}^\infty  f(x,b)\,dx\Big) (t-b) db  . \nonumber 
\end{eqnarray}
Denote
\begin{eqnarray}
\label{1.8a}
V(t)= \int_0^t d\tau \int_0^\tau  db \int_{-\infty}^\infty  v(x,b) dx ,\qquad F(t):=\int_0^t \Big( \int_{-\infty}^\infty  f(x,b)\,dx\Big) (t-b) db \,,
\end{eqnarray}
then (\ref{1.8}) and (\ref{1.8a}) imply
\[
V_{tt}-M^2 V=F,\qquad V(0)=V_t(0)=0\,.
\]
We easily find
\begin{eqnarray*}
V(t)
& = & 
\frac{1}{M} \int_0^t F(\tau ) \sinh (M(t-\tau ) )  d \tau   \,.
\end{eqnarray*}
Then (\ref{1.8}) implies
\begin{eqnarray*} 
\int_{-\infty}^\infty v(x,t)dx  
& = &
\int_0^t \Big( \int_{-\infty}^\infty  f(x,b)\,dx\Big) (t-b) db +M \int_0^t F(\tau ) \sinh (M(t-\tau ) )  d \tau  \\
& = &
\int_0^t \Big( \int_{-\infty}^\infty  f(x,b)\,dx\Big) (t-b) db \\
&  &
+M \int_0^t db 
\Big( \int_{-\infty}^\infty  f(x,b)\,dx\Big)
\int_b^t  d \tau  (\tau -b) \sinh (M(t-\tau ) ).   
\end{eqnarray*}
On the other hand
\[
\int_b^t  d \tau  (\tau -b) \sinh (M(t-\tau ) )   = -\frac{1}{M}(t-b)+ \frac{1}{M^2} \sinh (M(t-b ) )  
\]
implies
\begin{eqnarray*}
&&
\int_{-\infty}^\infty v(x,t)dx\\
& = &
\int_0^t \Big( \int_{-\infty}^\infty  f(x,b)\,dx\Big) (t-b) db\\
&  &
  + M 
\int_0^t db \Big( \int_{-\infty}^\infty  f(x,b)\,dx\Big)\Big( -\frac{1}{M}(t-b)+ \frac{1}{M^2} \sinh (M(t-b ) )  ) \Big)  \\
& = &  
\int_0^t  \Big( \int_{-\infty}^\infty  f(x,b)\,dx\Big)\Big( \frac{1}{M} \sinh (M(t-b ) ) \Big)  db\,.
\end{eqnarray*}
 Thus, for the arbitrary function $f \in C^\infty ({\mathbb R}\times [0,\infty))$ for all $t$ due to (\ref{3.11}) one has 
\begin{eqnarray*}
&  &
\int_0^t  \Big( \int_{-\infty}^\infty  f(x,b)\,dx\Big)\Big( \frac{1}{M} \sinh (M(t-b ) ) \Big)  db \\
& = &
 \int_{ 0}^{t} db \Big( \int_{-\infty}^\infty  f(x,b)\,dx \Big)  
\int_{   - (e^{-b}- e^{-t})}^{  e^{-b}- e^{-t}}  dz\,   
(4e^{-b-t })^{-M} \Big((e^{-t }+e^{-b})^2 - z^2\Big)^{-\frac{1}{2}+M    } \\
&  &
\hspace{2.5cm}\times F\Big(\frac{1}{2}-M   ,\frac{1}{2}-M  ;1; 
\frac{ ( e^{-b}-e^{-t })^2 -z^2 }{( e^{-b}+e^{-t })^2 -z^2 } \Big)   \,.
\end{eqnarray*}
It follows (\ref{3.9}).
Thus (i) is proved.

To prove case (ii) with $n$ is odd,  $n=2m+1$, $m  \in {\mathbb N}$, we use the identity
\[
1  = 
  \frac{\partial }{\partial r} 
\Big( \frac{1}{r} \frac{\partial }{\partial r}\Big)^{\frac{n-3}{2} } 
\frac{r^{n-2}}{\omega_{n-1} c_0^{(n)} } \int_{S^{n-1} } 
\, dS_y   
\]
and  take into consideration that the kernel 
\[ 
(4e^{-b-t})^{-M}
\left( (e^{-t}  + e^{-b} )^2 - r_1^2   \right)^{-\frac{1}{2}+M}
F\left(\frac{1}{2}-M,\frac{1}{2}-M;1; 
\frac{ (e^{-b}- e^{-t})^2-r_1^2}
{  (e^{-b}+ e^{-t})^2-r_1^2} \right)\!\! 
\]
is an even function of $r_1$.
In the case of (iii) when $n$ is even,  $n=2m$, $m  \in {\mathbb N}$, we apply the identity
\[
1  = 
 \frac{\partial }{\partial r} 
\Big( \frac{1}{r} \frac{\partial }{\partial r}\Big)^{\frac{n-2}{2} } 
\frac{2r^{n-2}}{\omega_{n-1} c_0^{(n)} } \int_{B_1^{n}(0)} \frac{1 }{\sqrt{1-|y|^2}} \, dV_y \,.
\]
The proposition is proven. 
\hfill $\Box$
\medskip

If we set in the above integrals $b=0$ then we get integral representations of the function 
$\sinh (Mt)$ depending on parameter $M>0$. We also note that   both sides of these formulas are translation invariant in $t$.
By passing to the limit as $M \to 0$ we arrive at the following corollary.
\begin{corollary} 
\label{CLt_tau}
The function $t-b$ with $t\ge b \ge 0$, 
can be represented as follows:\\
\mbox{\rm (i)} The one-dimensional integral  
\begin{eqnarray*} 
t-b 
& = &
  \int_{  - (e^{-b}- e^{-t})}^{e^{-b}- e^{-t}} 
\Big((e^{-t }+e^{-b})^2 - z^2\Big)^{-\frac{1}{2}}  F\Big(\frac{1}{2} ,\frac{1}{2};1; 
\frac{ ( e^{-b}-e^{-t })^2 -z^2 }{( e^{-b}+e^{-t })^2 -z^2 } \Big) \,   dz ;  \nonumber 
\end{eqnarray*}
\mbox{\rm (ii)} If $n$ is odd,  $n=2m+1$, $m  \in {\mathbb N}$, then with $c_0^{(n)} =1\cdot 3\cdot \ldots \cdot (n-2 )$, 
\begin{eqnarray*}
t-b 
& = &
2
  \int_{ 0}^{ e^{-b}- e^{-t}} dr_1 \,  \left(  \frac{\partial }{\partial r} 
\Big(  \frac{1}{r} \frac{\partial }{\partial r}\Big)^{\frac{n-3}{2} } 
\frac{r^{n-2}}{\omega_{n-1} c_0^{(n)} }  \!\!\int_{S^{n-1} } \, dS_y  
\right)_{r=r_1}   \nonumber \\
  &  & 
\times 
\left( (e^{-t}  + e^{-b} )^2 - r_1^2   \right)^{-\frac{1}{2}}
F\left(\frac{1}{2},\frac{1}{2};1; 
\frac{ (e^{-b}- e^{-t})^2-r_1^2}
{  (e^{-b}+ e^{-t})^2-r_1^2} \right); 
\end{eqnarray*} 
\mbox{\rm (iii)} If $n$ is even,  $n=2m$, $m  \in {\mathbb N}$, 
then with $c_0^{(n)} =1\cdot 3\cdot \ldots \cdot (n-1 )$, 
\begin{eqnarray*}
t-b 
& = &
2\int_{ 0}^{ e^{-b}- e^{-t}} dr_1 \,  \left( \frac{\partial }{\partial r} 
\Big( \frac{1}{r} \frac{\partial }{\partial r}\Big)^{\frac{n-2}{2} } 
\frac{2r^{n-1}}{\omega_{n-1} c_0^{(n)} }  \!\!\int_{B_1^{n}(0)} \frac{1 }{\sqrt{1-|y|^2}} \, dV_y 
\right)_{r=r_1}     \\
  &  & 
\times  
\left( (e^{-t}  + e^{-b} )^2 - r_1^2   \right)^{-\frac{1}{2}}
F\left(\frac{1}{2},\frac{1}{2};1; 
\frac{ (e^{-b}- e^{-t})^2-r_1^2}
{  (e^{-b}+ e^{-t})^2-r_1^2} \right)\!\! .  
\end{eqnarray*} 
\end{corollary}
\medskip

\section{The second order differential inequalities} 
\label{S_diffineq}

The second order differential inequality with the power decreasing kernel play key role in   proving
blow-up of the solutions of the semilinear equations. Kato's lemma \cite{Kato1980} 
allows us to derive from  inequality 
\[
\ddot w \geq b t^{-1-p} w^p,\qquad p>1, \,\, b>0, \quad t \, \,\,\mbox{\rm large}
\]
a boundedness  of the life-span
of solution with property $w_t \geq a >0$.   For the equation in de~Sitter spacetime the kernel of the
corresponding ordinary differential inequality decreases exponentially: 
\[
\ddot w \geq b e^{-Mt} w^p,\qquad p>1, \,\, b>0, \,M>0,\quad t \,\,\, \mbox{\rm large}.
\]
There is  a global  solution to the last  inequality. Hence,   
in order to reach exact conditions on  the 
involving functions we have to  generalize Kato's lemma. It is done in two following lemmas.
\begin{lemma} 
\label{L2.5}
Suppose $F(t) \in C^2([a,b))$, and 
\beqst
&  &
F(t) \ge  0\,, \qquad  \dot F(t) \geq 0 
\,, \qquad \ddot F(t) \ge  \Gamma (t)  F(t)^p \quad \mbox{\rm for all} \quad t \in [a,b)\,, 
\eeqst
where  $ \Gamma   \in C^1([a,\infty))$ is non-negative  function, $\Gamma (a) >0$,  and $p>1$. Assume that 
  for all  \, $t \in [a,b)$  \, either
\beqst
&  &
\dot \Gamma (t) \leq 0  \quad \mbox{\rm or } \quad  \Gamma(t) \geq const > 0  \,.
\eeqst
If there exists  \,  $ a_1 \in (a,b)$ \, such that 
\begin{equation}
\label{2.29} 
\frac{1}{\sqrt{p+1}} \int_a^{a_1} \Gamma   (s)^{1/2} ds  > \frac{\sqrt{2}}{p-1}F (a)^{(1- p )/2} \,, \quad \dot F(a)^2   \geq  \frac{2 }{p+1}    \Gamma   (a)F(a)^{ p+1 },
\end{equation}
then $b$ must be finite unless $\lim_{t \to \infty} F(t)  $ is finite.   
\end{lemma} 
\medskip

\noindent
{\bf Proof.}  First we consider the case of $ \dot \Gamma \leq 0$. The conditions of the lemma imply that 
derivative of the energy density function
is non-negative, 
\[
\frac{d}{dt} \left(F_t(t)^2  - \frac{2}{p+1} \Gamma   (t)  F(t)^{p+1} \right) \geq  0\quad \mbox{\rm for all} \quad t \in [a,b)\,.
\] 
We integrate the last inequality and obtain
\[
F_t(t)^2    \geq  \frac{2}{p+1} \Gamma   (t)   F(t)^{p+1} + F_t(a)^2  -\frac{2}{p+1} \Gamma   (a)   
F(a)^{p+1}\quad \mbox{\rm for all} \quad t \in [a ,b)\,.
\] 
In fact, according  to the second inequality of the condition (\ref{2.29}) we have 
\[
F_t(t)^2    \geq  \frac{2}{p+1} \Gamma   (t)   F(t)^{p+1} \quad \mbox{\rm for all} \quad t \in [a ,b)\,.
\] 
Hence,
\[
F_t(t)\geq \sqrt{\frac{2}{p+1}}\Gamma   (t)^{1/2} F(t)^{(p+1)/2}    \quad \mbox{\rm for all} \quad t \in [a ,b)\,.
\]
It follows
\[
\frac{d}{dt} \left( \frac{2}{1-  p }F^{1- (p+1)/2} (t) \right) \geq \sqrt{\frac{2}{p+1}}\Gamma   (t)^{1/2}     \quad \mbox{\rm for all} \quad t \in [a ,b)\,.
\]
Consequently,
\[
\frac{2}{1-  p }F^{(1- p )/2} (t) - \frac{2}{1-  p }F^{(1- p )/2} (a) \geq \sqrt{\frac{2}{p+1}}\int_a^t \Gamma   (s)^{1/2} ds    \,.
\]
According to the first inequality of the condition (\ref{2.29})   there exists  $a_1 > a $ such that
\begin{eqnarray*}
&  &
\frac{2}{1-  p }F^{(1- p )/2} (t) \\
& \geq &
\sqrt{\frac{2}{p+1}}\int_{a_1}^t \Gamma   (s)^{1/2} ds + 
\sqrt{\frac{2}{p+1}}\int_a^{a_1} \Gamma   (s)^{1/2} ds - \frac{2}{p-  1 }F^{(1- p )/2} (a)  \\
& \geq &
\sqrt{\frac{2}{p+1}}\int_{a_1}^t \Gamma   (s)^{1/2} ds 
\end{eqnarray*}
for all $t \in [a_1 ,b)$. Thus, 
for large $t$ we get contradiction. The case of uniformly positive function $\Gamma $ 
follows from Kato's Lemma~\cite{Kato1980}. Lemma is proven. 
\hfill $\Box$

  Next we turn to the case of the small energy and exponentially decreasing $\Gamma  (t ) $.

\begin{lemma} 
\label{L_ODIEXP}
Suppose $F(t) \in C^2([a,b))$, and 
\begin{equation}
\label{2.25c} 
F(t) \ge c_0 A(t) , \quad  F_{t}(t) \geq 0 , \quad 
F_{tt}(t) \ge \gamma  (t)A(t)^{- p } F(t)^p \quad \mbox{\rm for all} \,\, t \in [a,b) , 
\end{equation}
where  $A, \gamma  \in C^1([a,\infty))$ are non-negative  functions and $p>1$, $c_0>0$. Assume that  
\beq
\label{2.25a}
&  &
\lim_{t \to \infty} A(t) = \infty\,,
\eeq
  and that 
\beq
\label{2.25}
&  &
\frac{d}{dt} \left( \gamma  (t)A(t)^{- p }  \right) \leq 0\quad \mbox{\rm for all} \quad t \in [a,b)\,.
\eeq
If there exist $\varepsilon >0$ and $ c>0$ such that 
\begin{eqnarray}
\label{GammaAp}
& &
 \gamma  (t) \geq c A(t) (\ln A(t))^{2+\varepsilon}\quad \mbox{\rm for all} \quad t \in [a,b), 
\end{eqnarray}
then $b$ must be finite.
\end{lemma} 
\medskip

\noindent
{\bf Proof.}
There is a point $a_1 \geq a$ such that ${F}_{t}(a_1)>0$. Then $ F_{t}(t) \ge F_{t}(a_1)$ for all $t \geq a_1$ and consequently
\beqst
&  &
{F}(t) \ge \frac{1}{2}{F}_{t}(a_1) t \quad \mbox{\rm for all} \quad t \in [a_2,b)\,,
\eeqst
for sufficiently large $ a_2$. Furthermore, according to (\ref{2.25}) for the energy density  function we have
\[
\frac{d}{dt} \left(F_t(t)^2  -2\frac{1}{p+1} \gamma  (t)A(t)^{- p }  F(t)^{p+1} \right) \geq  0\quad \mbox{\rm for all} \quad t \in [a_1,b)\,.
\] 
The last inequality implies
\[
F_t(t)^2    \geq 2\frac{1}{p+1} \gamma  (t)A(t)^{- p }  F(t)^{p+1} + F_t(a_1)^2  -2\frac{1}{p+1} \gamma  (a_1)A(a_1)^{- p }  F(a_1)^{p+1}
\] 
for all \,$t \in [a_1,b)$. For sufficiently large $a_2\geq a_1$ using conditions (\ref{2.25c}), (\ref{2.25a}), and  (\ref{GammaAp})  we derive 
\beqst
\frac{1}{p+1} \gamma  (t)A(t)^{- p }  F(t)^{p+1} 
&  \geq &
\frac{1}{p+1} c_0^{p+1}\gamma  (t)A(t)  \\
& \geq &
 \frac{1}{p+1} cc_0^{p+1} A(t)^2 (\ln A(t))^{2+\varepsilon}\\
& \geq &
 F_t(a_1)^2   -2\frac{1}{p+1} \gamma  (a_1)A(a_1)^{- p }  F(a_1)^{p+1}  
\eeqst 
for all \,$t \in [a_2,b)$. Hence,
\[
F_t(t)\geq \sqrt{\frac{1}{p+1}}\gamma  (t)^{1/2} A(t)^{- p/2 }F(t)^{(p+1)/2}    \quad \mbox{\rm for all} \quad t \in [a_2,b)\,.
\]
It follows
\[
F_t(t)\geq \delta \gamma  (t)^{1/2} A(t)^{- p/2 } F(t)^{(p-1)/2  } (\ln F(t))^{-1-\varepsilon/2}  F(t) (\ln F(t))^{1+\varepsilon/2}  
\]
for all \,$t \in [a_2,b)$. But with sufficiently large $a_2 \geq a_1$ we obtain 
\[
 F(t)^{(p-1)/2  } (\ln F(t))^{-1-\varepsilon/2} \geq  \delta A(t)^{(p-1)/2  } (\ln A(t))^{-1-\varepsilon/2} 
\quad \mbox{\rm for all} \quad t \in [a_2,b)\,.
\]
Hence,
\[
F_t(t)\geq \delta \left( \gamma  (t) A(t)^{- 1  }  (\ln A(t))^{-2-\varepsilon} \right)^{1/2}   F(t) (\ln F(t))^{1+\varepsilon/2} 
  \quad \mbox{\rm for all} \quad t \in [a_2,b)
\]
implies
\[
F_t(t)\geq \delta  c   F(t) (\ln F(t))^{1+\varepsilon/2}   \quad \mbox{\rm for all} \quad t \in [a_2,b)\,.
\]
The last nonlinear differential inequality does not have   global solution with $F >0$. Lemma is proven. 
\hfill $\Box$

\begin{remark}
We note here that the equation
\beqst 
\ddot F(t) =   e^{-dt} F(t)^p\,,\quad d>0,   
\eeqst
has a global solution $F(t)=  c_F e^{\frac{d}{p-1}t}$, where $c_F= \left( {d}/(p-1) \right)^{2/(p-1)}$, while   
corresponding $A(t)=c_A e^{at}$, $a>0$,
and $\gamma (t)= c_\gamma  e^{(pa-d) t}$.  The condition (\ref{GammaAp}) implies   $a> d/(p-1)$. 
On the other hand, the first inequality of (\ref{2.25c}) holds only if  $a\leq  d/(p-1)$. 
 \end{remark}

\section{Nonexistence of the global solution for the integral equation associated with
the  Klein-Gordon equation}
\label{S_Proofs}

Since $G$ is a fundamental solution of the strictly hyperbolic operator, 
for every given function  
$u_0  \in C([0,T];$ $ L^q({\mb R}^n))\cap C^\infty ([0,T]\times {\mb R}^n )$ there exist $T_0 >0$ and solution $u  \in C([0,T_0]; L^q({\mb R}^n))$.
Moreover, for every given 
 $T$ one can prove existence of the solution
$u  \in C([0,T]; L^q({\mb R}^n))$ provided that $\sup_{t \in [0,T]} \| u_0 (\cdot ,t)\|_{L^q({\mb R}^n)}  $ is small enough.
Theorem~\ref{T2.1}  shows that the set of such  $T$, in general, is bounded.
\smallskip

\noindent
{\sl Proof of Theorem~\ref{T2.1}.} Let 
$u_0 \in C^\infty([0,\infty)\times {\mb R}^n )$ be a function with the permanently bounded support, that is
supp$\,u_0(\cdot,t) \subset \{\, x \in {\mathbb R}^n\, ;\, |x| \le constant\,\}$ for all $ t \geq 0$.
We denote $ \varphi _0 (x):= u_0(x,0) $ and $ \varphi _1 (x):= \partial_t u_0(x,0) $. One can find 
$u_0 $ such that 
\begin{equation}
\label{2.9}
\int_{{\mathbb R}^n} u_0(x,t)  dx = C_0 \cosh (Mt) +C_1\frac{1}{M} \sinh(M t) 
\qquad \mbox{\rm for all} \quad t \geq 0\, , 
\end{equation}
where
\begin{equation}
\label{2.10}
C_0:= \int_{{\mathbb R}^n} \varphi _0(x )   dx,  \quad
C_1:= \int_{{\mathbb R}^n } \varphi _1(x )   dx \,.
\end{equation}
 The solution  of the problem (\ref{3.5}) 
with the data $\varphi _0 (x)$,  $ \varphi _1 (x) \in C^\infty_0(  {\mb R}^n )$
exemplifies such function.
Indeed, this unique smooth solution obeys finite propagation speed property that implies 
supp$\,u_0(\cdot,t) \subset \{\, x \in {\mathbb R}^n\, ;\, |x| \le R_0+ 1-e^{-t} \leq R_0+ 1\,\}$ 
if supp$\,\varphi _0$, supp$\,\varphi _1   \subset \{\, x \in {\mathbb R}^n\, ;\, |x| \le R_0\,\}$.   
In order to check (\ref{2.9}) for that solution $u_0$ we  integrate (\ref{3.5}) with respect to $x$ over $ {\mathbb R}^n$ and 
then solve the initial problem with data  (\ref{2.10})  for the obtained ordinary differential equation.
\smallskip

Suppose that $u  \in C([0,\infty); L^q({\mb R}^n))$ with permanently bounded support is a solution to 
(\ref{2.7}) generated by $u_0 $. 
According to the definition of the solution, for every given $T>0$ we have 
\[
G\left[\Gamma (\cdot ) \left( \int_{{\mb R}^n} |u (y,\cdot )|^p dy \right)^\beta  |u |^p\right]  \in C ([0,T]; L^{q }({\mb R}^n))
\]
and $
  u (x,0)  =   \varphi _0(x)\,, 
\quad  u_t (x,0) =  \varphi _1(x)\,.
$
Then   $u  \in C([0,\infty); L^1({\mb R}^n))$ and 
we can integrate equation (\ref{2.7}):
\begin{equation}
\label{1.7a} 
\int_{{\mb R}^n} u(x,t)  \,dx = \int_{{\mb R}^n} u_0(x,t)  \,dx 
+ \int_{{\mb R}^n } G\left[\Gamma (\cdot ) \left( \int_{{\mb R}^n} |u (y,\cdot )|^p dy \right)^\beta  |u |^p\right] (x,t)  \,dx   .
\end{equation}
In particular,
\begin{eqnarray*}
& &
 \int_{{\mb R}^n} u (x,0) \,dx= \int_{{\mb R}^n} \varphi _0(x)\,dx, 
\quad \int_{{\mb R}^n} u_t (x,0)\,dx = \int_{{\mb R}^n} \varphi _1(x)\,dx\,.
\end{eqnarray*}
To evaluate the last term  of (\ref{1.7a}) we apply Proposition~\ref{Lt_tau}. Consider the case of odd $n \geq 3$. Then, 
for the smooth function $u=u(x,t) $     we obtain 
 \begin{eqnarray*} 
& &
\int_{{\mathbb R}^n} G[\Gamma (\cdot )|u|^p ](x,t) \, dx  \\
 & = &
\int_{{\mathbb R}^n} \, dx 2\int_{ 0}^{t} db
  \int_{ 0}^{ e^{-b}- e^{-t}} dr_1     \Big(  \frac{\partial }{\partial r} 
\Big(  \frac{1}{r} \frac{\partial }{\partial r}\Big)^{\frac{n-3}{2} } 
\frac{r^{n-2}}{\omega_{n-1} c_0^{(n)} }  \\
&  &
\times \int_{S^{n-1} } \Big[ \Gamma  (b ) \Big( \int_{{\mathbb R}^n}  |u(z,b )|^{p}  dz \Big)^\beta|u (x+ry,b) |^{p} \Big]\, dS_y  
\Big)_{r=r_1}    \\
  &  & 
\times  (4e^{-b-t})^{-M}
\left( (e^{-t}  + e^{-b} )^2 - r_1^2   \right)^{-\frac{1}{2}+M}\\
  &  & 
\times 
F\left(\frac{1}{2}-M,\frac{1}{2}-M;1; 
\frac{ (e^{-b}- e^{-t})^2-r_1^2}
{  (e^{-b}+ e^{-t})^2-r_1^2} \right)\,. 
\end{eqnarray*}
Therefore,
\begin{eqnarray*} 
& &
\int_{{\mathbb R}^n} G[\Gamma (\cdot )|u|^p ](x,t) \, dx  \\
& = &
2\int_{ 0}^{t} db
  \int_{ 0}^{ e^{-b}- e^{-t}} dr_1  \Big(  \frac{\partial }{\partial r} 
\Big(  \frac{1}{r} \frac{\partial }{\partial r}\Big)^{\frac{n-3}{2} } 
\frac{r^{n-2}}{\omega_{n-1} c_0^{(n)} }  \\
&  &
\times \int_{S^{n-1} } \Big[ \Gamma  (b ) \Big( \int_{{\mathbb R}^n}  |u(z,b )|^{p}  dz \Big)^\beta
\left( \int_{{\mathbb R}^n}  |u (x+ry,b) |^{p}\, dx \right) \Big]\, dS_y  
\Big)_{r=r_1}   \nonumber \\
  &  & 
\times  (4e^{-b-t})^{-M}
\left( (e^{-t}  + e^{-b} )^2 - r_1^2   \right)^{-\frac{1}{2}+M}\\
&  &
\times 
F\left(\frac{1}{2}-M,\frac{1}{2}-M;1; 
\frac{ (e^{-b}- e^{-t})^2-r_1^2}
{  (e^{-b}+ e^{-t})^2-r_1^2} \right)   
\end{eqnarray*}
implies,
\begin{eqnarray*}  
\int_{{\mathbb R}^n} G[\Gamma (\cdot )|u|^p ](x,t) \, dx   
& = &
2\int_{ 0}^{t} db
  \int_{ 0}^{ e^{-b}- e^{-t}} dr_1 \,  \Big(  \frac{\partial }{\partial r} 
\Big(  \frac{1}{r} \frac{\partial }{\partial r}\Big)^{\frac{n-3}{2} } 
\frac{r^{n-2}}{\omega_{n-1} c_0^{(n)} }  \\
&  &
\times \int_{S^{n-1} } \Big[ \Gamma  (b ) \Big( \int_{{\mathbb R}^n}  |u(z,b )|^{p}  dz \Big)^{\beta+1}
\Big]\, dS_y  
\Big)_{r=r_1}   \nonumber \\
  &  & 
\times  (4e^{-b-t})^{-M}
\left( (e^{-t}  + e^{-b} )^2 - r_1^2   \right)^{-\frac{1}{2}+M} \\
&  &
\times 
F\left(\frac{1}{2}-M,\frac{1}{2}-M;1; 
\frac{ (e^{-b}- e^{-t})^2-r_1^2}
{  (e^{-b}+ e^{-t})^2-r_1^2} \right)\,.
 \end{eqnarray*} 
We obtain,
 \begin{eqnarray*} 
\int_{{\mathbb R}^n} G[\Gamma (\cdot )|u|^p ](x,t)  dx  
  & = &
\int_{ 0}^{t} db\Big[ \Gamma  (b ) \Big( \int_{{\mathbb R}^n}  |u(z,b )|^{p}  dz \Big)^{\beta+1}
\Big]\int_{ 0}^{ e^{-b}- e^{-t}} dr_1 \\
&  &
\times 
  2   \left(  \frac{\partial }{\partial r} 
\Big(  \frac{1}{r} \frac{\partial }{\partial r}\Big)^{\frac{n-3}{2} } 
\frac{r^{n-2}}{\omega_{n-1} c_0^{(n)} }  \!\!\int_{S^{n-1} } \, dS_y  
\right)_{r=r_1}   \\
  &  & 
\times (4e^{-b-t})^{-M}
\left( (e^{-t}  + e^{-b} )^2 - r_1^2   \right)^{-\frac{1}{2}+M} \\
&  &
\times 
F\left(\frac{1}{2}-M,\frac{1}{2}-M;1; 
\frac{ (e^{-b}- e^{-t})^2-r_1^2}
{  (e^{-b}+ e^{-t})^2-r_1^2} \right) \\
& = &
\int_{ 0}^{t}  \Gamma  (b ) \Big( \int_{{\mathbb R}^n}  |u(z,b )|^{p}  dz \Big)^{\beta+1} 
 \frac{1}{M} \sinh (M(t-b ) ) \,  db .
\end{eqnarray*} 
Thus, for the solution  $u=u(x,t) $ we have proven
\begin{eqnarray*} 
\int_{{\mathbb R}^n} G[\Gamma (\cdot )|u|^p ](x,t) \, dx 
& = &
\int_{ 0}^{t}  \Gamma  (b ) \Big( \int_{{\mathbb R}^n}  |u(z,b )|^{p}  dz \Big)^{\beta+1} 
 \frac{1}{M} \sinh (M(t-b ) ) \,  db \, .
\end{eqnarray*} 
Hence (\ref{1.7a}) reads
\begin{eqnarray*} 
&  &
\int_{{\mathbb R}^n} u(x,t)  \,dx \\
& = &
\int_{{\mathbb R}^n} u_0(x,t)  \,dx 
+  \int_{ 0}^{t}  \Gamma  (b ) \Big( \int_{{\mathbb R}^n}  |u(z,b )|^{p}  dz \Big)^{\beta+1} 
 \frac{1}{M} \sinh (M(t-b ) ) \,  db \, .  
\end{eqnarray*}
Taking into account (\ref{2.9}) and  (\ref{2.10}) we derive
\begin{eqnarray*}
\int_{{\mathbb R}^n} u(x,t)  \,dx 
& = &
\frac{1}{2}\left(C_0+\frac{C_1}{M} \right)e^{M t} +\frac{1}{2}\left(C_0-\frac{C_1}{M} \right) e^{-M t}  \nonumber\\ 
&  &
+  \int_{ 0}^{t}  \Gamma  (b ) \Big( \int_{{\mathbb R}^n}  |u(z,b )|^{p}  dz \Big)^{\beta+1} 
 \frac{1}{M} \sinh (M(t-b ) ) \,  db   \, .
\end{eqnarray*}
We discuss separately   two  cases:  with  positive curved mass, $M>0$, and vanishing curved mass, $M=0$, respectively.
\medskip

\noindent
In the case of $M>0$ we obtain 
\begin{eqnarray*}
 \int_{{\mathbb R}^n} u(x,t)  \,dx 
& =  &
   C_0 \cosh(Mt)+ \frac{C_1}{M}  \sinh (Mt)   \\
& +   &
\int_{ 0}^{t}  \Gamma  (b ) \Big( \int_{{\mathbb R}^n}  |u(z,b )|^{p}  dz \Big)^{\beta+1} 
 \frac{1}{M} \sinh (M(t-b ) ) \,  db     \,.
\end{eqnarray*}
Denote 
\begin{eqnarray*}
F(t)
& := &
  \int_{{\mathbb R}^n }  u(x,t)   \,dx     \,,
\end{eqnarray*}
then the function $F(t)$ is
\begin{eqnarray*}
F(t)
& = &
   C_0 \cosh(Mt)+ \frac{C_1}{M}  \sinh (Mt)   \\
&  &
+  \int_{ 0}^{t}  \Gamma  (b ) \Big( \int_{{\mathbb R}^n}  |u(z,b )|^{p}  dz \Big)^{\beta+1} 
 \frac{1}{M} \sinh (M(t-b ) ) \,  db    \,.
\end{eqnarray*}
It follows $ F \in C^2([0,\infty))$. More precisely, 
\begin{eqnarray}
\label{1.17a}
\dot F(t) 
& = &
   C_1\cosh(Mt)+ MC_0  \sinh (Mt)    \\
&  & 
+   \int_{ 0}^{t}  \Gamma  (b ) \Big( \int_{{\mathbb R}^n}  |u(z,b )|^{p}  dz \Big)^{\beta+1} 
\cosh (M(t-b ) ) \,  db   \,, \nonumber \\ 
\ddot F(t) 
& = &
 M^2F(t)    +    \Gamma  (t ) \Big( \int_{{\mathbb R}^n}  |u(z,t )|^{p}  dz \Big)^{\beta+1}    \,.
\end{eqnarray}
In particular, since $ \Gamma  (t )\geq 0$, we obtain
\begin{equation}
\label{3.21new}
F(t)
\geq 
  C_0 \cosh(Mt)+ \frac{C_1}{M}  \sinh (Mt)  
\,\, \mbox{\rm and} \,\, \ddot F(t)  
 \geq 
  \Gamma  (t ) \Big( \int_{{\mathbb R}^n}  |u(z,t )|^{p}  dz \Big)^{\beta+1}  .
\end{equation}
On the other hand, since the  solution $u=u(x,t)$ has permanently bounded support, then  
supp$\,u(\cdot,t) \subset \{\, x \in {\mathbb R}^n\, ;\, |x| \le R \,\}$ for some positive number $R$.  Using the compact support of $u(\cdot,t)$ and H\"older's inequality we get
with \,  $\tau_n$  the volume of the
unit ball in ${\mathbb R}^n$, 
\begin{eqnarray*}
\left| \int_{{\mathbb R}^n} u (x,t) \,dx \right|^{p} 
& \le &
\left( \int_{|x| \le R} 1 \,dx \right)^{ p-1 } 
\left( \int_{|x| \le R} |u (x,t)|^{p} \,dx \right)\\
& = &
\tau_n \Gamma (t)^{-1/(\beta+1)} R^{n (p-1)  }
\left( \Gamma (t)^{1/(\beta+1)}  \int_{{\mathbb R}^n} |u (x,t)|^{p} \,dx \right)\\
& = &
 \tau_n  \Gamma (t)^{-1/(\beta+1)}  R^{n (p-1)  }\Big(\ddot{F}(t)-  M^2F(t) \Big)^{1/(\beta+1)} \\
& \leq  &
 \tau_n  \Gamma (t)^{-1/(\beta+1)}  R^{n (p-1)  }\ddot{F}(t)^{1/(\beta+1)}  \,.
\end{eqnarray*}
Here we assume $ \Gamma  (t ) > 0$. Thus
\[
\ddot{F}(t) 
\ge  \tau_n^{-(\beta +1)}R^{-n(p-1)  (\beta +1)} \Gamma (t) |F(t)|^{p(\beta +1)}
\quad \mbox{\rm for all} \quad t\in   [0,\infty)\,.
\]
By means of the inequality 
\[
M C_0  +  C_1   >0
\]
 we   conclude that $F(t)\geq 0 $ and that 
\[
\ddot{F}(t) 
\ge  \delta _0   \Gamma (t)  F(t) ^{p(\beta +1)}\qquad 
\mbox{\rm for large} \,\, \, t \quad \mbox{\rm with} \,\, \,\delta _0  >0 \,.
\]
Hence, for appropriate $C_0$ and $C_1$   the last inequality together with (\ref{1.17a}) to (\ref{3.21new})
 implies the following system of the ordinary differential inequalities
\[ 
  \left\{ 
\begin{array}{ccccc}
\dsp F(t) & \geq  & \hspace{- 0.4cm}  C_0 \cosh(Mt)+ \frac{C_1}{M}  \sinh (Mt)   \quad & \mbox{\rm for all}&  \quad t \in [a,b),\\
\dsp \dot F(t) & \geq &   C_1\cosh(Mt)+ MC_0  \sinh (Mt)  \quad & \mbox{\rm for all} & \quad t \in [a,b),\\
\dsp \ddot F(t)  & \geq & \hspace{-1.4cm} \delta _0   \Gamma (t)  F(t) ^{p(\beta +1)}\quad & \mbox{\rm for all} & \quad t \in [a,b).   
 \end{array} \right. 
\] 
The Lemma~\ref{L2.5} shows that if $F(t) \in C^2([0,b)) $ and the energy of particle is large, then $b$ must be finite.  

The conditions of the Lemma~\ref{L2.5} are fulfilled on $(0,b)$ for the function
\[
\Gamma (t)=\delta _0 e^{\gamma t}, \quad \gamma \in {\mathbb R},
\]
with $\gamma >0$ without any condition on the energy. They are fulfilled with $\gamma <0$ if the kinetic energy and the potential energy are 
sufficiently large, that is $C_0>0$, $C_1>0$, and  
\[
C_1 \geq \sqrt{\frac{ 2\delta _0 }{p+1}}C_0^{(p+1)/2}   \quad \mbox{\rm and} \quad C_0^{p-1} > \frac{\gamma ^2(p+1)}{ \delta _0 (p-1) }\,.
\]

  Next we turn to the case of the small energy and exponentially decreasing $\Gamma  (t ) $.
We apply Lemma~\ref{L_ODIEXP} with $A(t)= e^{Mt}$ and $p$ replaced with  $p(\beta +1)$. More precisely, if we set 
\[
A(t) = e^{Mt},\qquad \gamma (t) = \Gamma (t)e^{Mp(\beta +1) t},
\]
then the conditions of the last lemma read: 
\[
p(\beta +1)>1 \quad \mbox{\rm and } \quad  \Gamma_t   (t)   \leq 0 \quad \mbox{\rm for all } \quad t \in [0,\infty).
\]
The last inequality follows from the monotonicity of $\Gamma (t)$. By the condition of the theorem, there exist $\varepsilon >0$ and $ c>0$ such that 
\begin{eqnarray*}
& &
\Gamma (t) \geq c e^{-M(p(\beta +1)-1 )t} t^{2+\varepsilon}\quad \mbox{\rm for all} \quad t \in [a,b), 
\end{eqnarray*} 
that coincides with (\ref{GammaAp}). The case of $M>0$ is proved.
\medskip

\noindent
Now consider the case of $M=0$.  
Let 
\begin{eqnarray*}
C_0:= \int_{{\mathbb R}^n} \varphi _0(x )   dx  ,  \quad
C_1:= \int_{{\mathbb R}^n}  \varphi _1(x )   dx  ,  
\qquad  C_1 >0.
\end{eqnarray*}
Then  Corollary~\ref{CLt_tau}   allows us to write 
\begin{eqnarray*} 
\int_{{\mathbb R}^n} G[\Gamma (\cdot )|u|^p ](x,t) \, dx 
& = &
\int_{ 0}^{t}  \Gamma  (b ) \Big( \int_{{\mathbb R}^n}  |u(z,b )|^{p}  dz \Big)^{\beta+1} 
(t-b ) \,  db .
\end{eqnarray*} 
Hence (\ref{1.7a}) reads:
\[
\int_{{\mathbb R}^n}  u(x,t)  \,dx = \int_{{\mathbb R}^n}  u_0(x,t)  \,dx 
+  \int_{ 0}^{t}  \Gamma  (b ) \Big( \int_{{\mathbb R}^n}  |u(z,b )|^{p}  dz \Big)^{\beta+1} \,.  
\]
Now we choose a function $u_0 \in C^\infty([0,\infty)\times {\mathbb R}^n)$ such that 
\[
\int_{{\mathbb R}^n}  u_0(x,t)  dx = C_0+C_1 t \, .
\]
The solution  of the problem (\ref{3.5}) with $M=0$ exemplifies such functions. Thus
\begin{eqnarray*}
\int_{{\mathbb R}^n}  u(x,t)  \,dx 
& = &
C_0+C_1t
+ \, 
 \int_{ 0}^{t}  \Gamma  (b ) \Big( \int_{{\mathbb R}^n}  |u(z,b )|^{p}  dz \Big)^{\beta+1} 
(t-b )\,  db \, .
\end{eqnarray*}
Denote 
\begin{eqnarray*}
F(t)
& := &
  \int_{{\mathbb R}^n}  u(x,t)  \,dx     \,,
\end{eqnarray*}
then 
\begin{eqnarray*}
F(t)
& = &
 C_0+ C_1 t   
+  \int_{ 0}^{t}  \Gamma  (b ) \Big( \int_{{\mathbb R}^n}  |u(z,b )|^{p}  dz \Big)^{\beta+1} (t-b )\,  db \,.
\end{eqnarray*}
It follows $F \in C^2([0,\infty))$. More precisely,
\begin{eqnarray}
\dot F(t) 
& = &
   C_1    
+   \int_{ 0}^{t}  \Gamma  (b ) \Big( \int_{{\mathbb R}^n}  |u(z,b )|^{p}  dz \Big)^{\beta+1} \,  db \,, \nonumber \\
\label{1.17}
\ddot F(t) 
& = &
\Gamma  (t ) \Big( \int_{{\mathbb R}^n}  |u(z,t )|^{p}  dz \Big)^{\beta+1}  \,.
\end{eqnarray}
In particular,
\begin{eqnarray}
\label{3.21}
F(t)
\geq 
 C_0+ C_1 t 
\quad \mbox{\rm and} \quad \ddot F(t)  
 = 
 \Gamma  (t ) \Big( \int_{{\mathbb R}^n}  |u(z,t )|^{p}  dz \Big)^{\beta+1}     \,.
\end{eqnarray}
On the other hand according to (\ref{1.17}) we obtain
\begin{eqnarray*}
\left| \int_{{\mathbb R}^n} u (x,t) \,dx \right|^{p} 
& \le &
\left( \int_{|x| \le R} 1 \,dx \right)^{ p-1  } 
\left( \int_{|x| \le R} |u (x,t)|^{p} \,dx \right)\\
& = &
\tau_n \Gamma (t)^{-1/(\beta+1)} R^{n (p-1)  }
\left( \Gamma (t)^{1/(\beta+1)}  \int_{{\mathbb R}^n} |u (x,t)|^{p} \,dx \right)\\
& \leq  &
 \tau_n  \Gamma (t)^{-1/(\beta+1)}  R^{n (p-1)   }\ddot{F}(t)^{1/(\beta+1)}  \,.
\end{eqnarray*}
Thus
\[
\ddot{F}(t) 
\ge  \tau_n^{-(\beta +1)}R^{-n(p-1)  (\beta +1)} \Gamma (t) |F(t)|^{p(\beta +1)}
\]
for all $t$ in  $[0,\infty)$. By means of the condition \,$C_1   >0$\,
 we   conclude
\[
\ddot{F}(t) 
\ge  C \Gamma (t)  F(t) ^{p(\beta +1)}\qquad 
\mbox{\rm for large} \,\,  t \quad \mbox{\rm with} \,\, C>0 \,.
\]
But for appropriate $C_0$ and $C_1$ one has $F(t) >0$ and the last inequality together with (\ref{3.21})
 implies 
 \begin{eqnarray*}
\left\{ \begin{array}{ccccc}
 \dsp F(t)& \geq &  \hspace{-1cm} C_0+ C_1 t \quad &\mbox{\rm for all} &  t \in [a,b),\\
\dsp \ddot F(t) & \geq & \delta _0  \Gamma (t)  F(t) ^{p(\beta +1)}\quad & \mbox{\rm for all}&  t \in [a,b).    
 \end{array} \right. 
  \end{eqnarray*}
The  Kato's Lemma~2~\cite{Kato1980} shows that if $F(t) \in C^2([0,b)) $ and $\Gamma (t) \geq t^{-1-p(\beta +1)} $ with $p(\beta +1) >1$,  then $b$ must be finite.
 Theorem is proven. \hfill $\square$
\begin{remark}
In fact, we have proved that any solution $u =u(x,t)$ with permanently bounded support blows up if either $MC_0+C_1>0$ and $M>0$ or $C_1>0$ and $M=0$.
\end{remark}

 \medskip

\noindent
{\sl Proof of Theorem~\ref{T2.1a}.} The case of $\gamma \geq 0$ is covered by Theorem~\ref{T2.1} and implies a blow-up even for the small data.
Therefore, we restrict ourselves to the case of $\gamma < 0$. Then,   
with a special choice of $C_0 $ and  $C_1$  after arguments have been used in the proof of Theorem~\ref{T2.1}
we arrive at the following system of the ordinary differential inequalities
 \begin{eqnarray*}
\left\{ \begin{array}{ccccc}
 \dsp F(t) & \geq &   \hspace{-1.6cm} Ce^{ Mt}   \quad & \mbox{\rm for all} &  t \in [0,b),\\
\dsp \dot F(t) & \geq &  \hspace{-1.6cm}   C e^{ Mt}   \quad & \mbox{\rm for all} & t \in [0,b),\\
\dsp \ddot F(t) & \geq &  \delta _0   e^{\gamma t} F(t) ^{p(\beta +1)}\quad & \mbox{\rm for all} & t \in [0,b),    
 \end{array} \right. 
  \end{eqnarray*}
where $C>0$ and $ \delta _0 >0 $. We claim that $b < \infty$. Indeed, we check conditions of Lemma~\ref{L2.5} with 
\[
\Gamma (t)= \delta _0   e^{\gamma t} \,. 
\]
The condition  (\ref{2.29}),
\begin{eqnarray*} 
\frac{1}{\sqrt{p+1}} \int_0^{a_1} \Gamma   (s)^{1/2} ds  > \frac{\sqrt{2}}{p-1}F^{(1- p )/2} (0) \,, 
\quad \dot F^2(0)   \geq  \frac{2 }{p+1}    \Gamma   (0)F(0)^{ p+1 },
\end{eqnarray*}
reads:
\begin{eqnarray*} 
\frac{1}{\sqrt{p+1}} \int_0^{a_1} \delta _0^{1/2}   e^{\gamma s/2}  ds  > \frac{\sqrt{2}}{p-1}C_0^{(1- p )/2}   \,, 
\quad   C_1^2    \geq  \frac{2 }{p+1}    \delta _0C_0^{ p+1 }.
\end{eqnarray*}
The first inequality is fulfilled if $C_0$, that is the initial potential energy,  is sufficiently large, while the second one is 
fulfilled if $C_1$, that is the initial kinetic energy,  is  large enough. Theorem is proven. \hfill $\square$

\end{document}